\documentclass[twocolumn]{aastex62}

\usepackage[nameinlink]{cleveref}

\usepackage{amsmath}
\usepackage{soul}

\newcommand*{\http}[1]{\href{http://#1}{#1}}

\shorttitle{Searching for Stellar Variability}
\shortauthors{Thomas and Kahn}

\begin{document}

\title{Searching for Sub-Second Stellar Variability with Wide-Field Star Trails and Deep Learning}

\author{David Thomas}
\affiliation{Institute for Computational and Mathematical Engineering, Stanford University, Stanford, CA 94305, USA}
\affiliation{Kavli Institute for Particle Astrophysics and Cosmology, Stanford University, Stanford, CA 94309, USA}
\altaffiliation{LSSTC Data Science Fellow}

\author{Steven M. Kahn}
\affiliation{Kavli Institute for Particle Astrophysics and Cosmology, Stanford University, Stanford, CA 94309, USA}
\affiliation{Department of Physics, Stanford University, Stanford CA, 94305, USA}
\affiliation{SLAC National Accelerator Laboratory, Menlo Park CA, 94025, USA}
\affiliation{LSST Project Office, 933 N. Cherry Ave, Tucson AZ, 85719, USA}



\begin{abstract}
We present a method that enables wide field ground-based telescopes to scan the sky for sub-second stellar variability. The method has operational and image processing components. The operational component is to take star trail images. Each trail serves as a light curve for its corresponding source and facilitates sub-exposure photometry. We train a deep neural network to identify stellar variability in wide-field star trail images. We use the Large Synoptic Survey Telescope (LSST) Photon Simulator to generate simulated star trail images and include transient bursts as a proxy for variability. The network identifies transient bursts on timescales down to 10 milliseconds. We argue that there are multiple fields of astrophysics that can be advanced by the unique combination of time resolution and observing throughput that our method offers.

\end{abstract}



\correspondingauthor{David Thomas}
\email{dthomas5@stanford.edu}

\section{Introduction} \label{sec:intro}

The universe remains relatively unexplored on short time scales in the optical region of the electromagnetic spectrum. Charge coupled devices (CCDs) have been the detectors of choice in astronomy for over four decades. These large pixel arrays provide high quantum efficiency with excellent spatial resolution, but are conventionally operated in integration mode with exposure durations of order tens of seconds. Such long exposures preclude these instruments from imaging astrophysics that manifests on shorter time scales. High-speed photometric surveys, which we define as imaging large areas of sky with time resolution below one second, have been mostly unavailable, both for the study of known variable sources and for the search for new phenomena.

The violent and rapidly varying radiation from black holes, neutron stars, and white dwarfs makes them promising targets for high time resolution imaging. The rotation, pulsation, and local accretion dynamics of these compact stellar remnants tends to occur on time scales ranging from seconds to milliseconds. Their extreme density also makes them an excellent testing ground for nuclear, quantum, and gravitational physics \citep{2004Sci...304..536L, 2016PhRvD..94h4002Y, 2017Natur.551...80K}.

Compact stellar remnants are one of many applications. High-speed optical photometry has also supplemented the study of brown dwarfs, cataclysmic variable stars, eclipsing binary stars, X-ray binary stars, extrasolar planets, flare stars, active galactic nuclei, asteroseismology, and atmospheres of solar system objects \citep{1988Natur.336..452H, 2007MNRAS.378..825D}. Combining high-speed imaging with a wide-field instrument opens up the possibility of serendipitously observing a Kuiper Belt occultation, the immediate afterglow of a gamma ray burst, an optical counterpart of a fast radio bursts, and other rare phenomena \citep{2013AJ....146...14Z}.

The majority of existing high-speed optical imaging instruments, such as ULTRACAM, ULTRASPEC, CHIMERA, and HiPERCAM take advantage of frame transfer and electron multiplying CCDs \citep{2007MNRAS.378..825D,2014MNRAS.444.4009D,2016SPIE.9908E..0YD, 2016MNRAS.457.3036H}. By limiting the camera readout to a small window surrounding a source of interest, these instruments can achieve high sample rates, even over 1,000 Hz.

In this work, we are primarily interested in developing the capacity to detect sub-second stellar variability over a \textit{wide field of view}. ULTRACAM for example, can image a $1024 \times 1024$ pixel array at 40 Hz.  Another promising approach for scanning the sky for sub-second variability is to operate CCDs in a continuous readout mode. \cite{2009AJ....138..568B} achieved 200 Hz photometry with continuous readout on MEGACAM.

We show how similar performance can be achieved at existing facilities through a relatively trivial modification to the observing plan. We revisit an idea originally introduced by \cite{1986PASP...98..802H}: that star trails provide sub-exposure time resolution. Almost every major telescope is capable of producing star trail images by simply turning off the tracking. The principle hurdle to this method is the need to process these unorthodox images. For this, we leverage deep learning.

Deep learning has achieved impressive results in many areas of machine learning. While neural networks and other paradigms in the field have a long history \citep{McCullochPitts43,Hochreiter:1997:LSM:265493.264179,lecun-gradientbased-learning-applied-1998,Hinton:2006:FLA:1161603.1161605,Bengio:2009:LDA:1658423.1658424}, increases in computational capacity, data sizes, and a series of practical tricks \citep{Glorot10understandingthe,DBLP:journals/corr/abs-1207-0580,journals/corr/KingmaB14,Ioffe:2015:BNA:3045118.3045167} have allowed deep neural networks to recently realize their full potential. From 2012 to 2014 deep learning methods broke through stubborn hurdles in object recognition \citep{NIPS2012_4824}, natural language processing \citep{Sutskever:2014:SSL:2969033.2969173}, and speech recognition \citep{graves2013speech}. In 2014, Ian Goodfellow introduced the powerful paradigm of generative adversarial networks \citep{NIPS2014_5423}. Iconic successes such as reaching human level performance in Atari games \citep{mnih2013playing,mnih-dqn-2015} and beating Go professionals \citep{SilverHuangEtAl16nature,silver2017mastering} also brought significant attention to these methods.

Astronomy is a data rich field and presents many opportunities for deep learning. \cite{2015MNRAS.450.1441D} used a convolutional neural network to predict galaxy morphologies in the Galaxy Zoo project. \cite{2017MNRAS.467L.110S} used a generative adversarial network to recover astrophysical features in images beyond the deconvolution limit. \cite{2018MNRAS.473.3895L} developed a method to find galaxy-galaxy strong lenses. \cite{2017Natur.548..555H} reduced the time to analyze strong gravitational lenses by seven orders of magnitude. \cite{2018AJ....155...94S} identified new exoplanets. \cite{2017arXiv171107966G} improved gravitational wave detection. \cite{2017arXiv170906257M} classified stellar light curves. Finally, \cite{2017arXiv171001422S} used deep learning for image subtraction, which is closest to our application. 

We design a deep neural network to sift through wide field star trail images and detect variability. The input to the network is a simulated star trail image and the output is an image containing only the excess flux from variability. We argue that this technique is well suited for the diverse aforementioned applications.

This paper is organized as follows. In Section \ref{sec:data} we describe the various LSST simulators and how we use them to produce training, development, and test datasets for our network. In Section \ref{sec:network} we describe our network architecture and training process. In Section \ref{sec:results} we assess the performance of our network with qualitative and quantitative measures. In Section \ref{sec:discussion} we review avenues for future work. In Section \ref{sec:conclusion} we summarize our findings.

\section{Data}
\label{sec:data}

We use simulated Large Synoptic Survey Telescope (LSST) images for our experiments. The LSST is a large ground-based telescope currently under construction, that will perform a ten-year imaging survey of the entire southern hemisphere sky, starting in October, 2022. The 8-m class primary mirror and 3.2 Gigapixel camera collectively provide a system etendue, or flux gathering capacity, of 319.5\ $\text{m}^2\text{deg}^2$, roughly ten times larger than that of any previous or planned survey facility \citep{2008arXiv0805.2366I,2009arXiv0912.0201L}. The LSST's unprecedented imaging volume highlights the full potential of our method. The LSST also has a rich simulation ecosystem \citep{2014SPIE.9150E..14C,2014SPIE.9150E..15D,2015ApJS..218...14P}. We use these tools to generate the data that our network is trained on.

\begin{figure*}[htb]
\center
\includegraphics{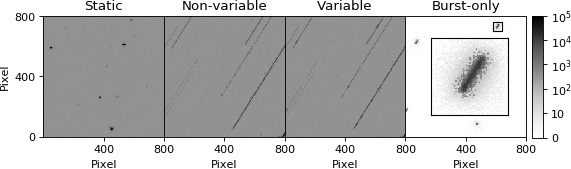}
\caption{The same catalog simulated with the four different simulation modes. From right to left: static, non-variable, variable, and burst-only. The images are $800 \times 800$ pixel crops from larger $4000 \times 4072$ pixel single CCD LSST images. The inset in the burst-only image shows a burst with higher resolution.}
\label{fig:1}
\end{figure*}

We produce simulated data in three stages. The first produces observing targets, the second produces corresponding catalogs, and the third produces corresponding images. For the first stage, we use the LSST Operations Simulator \citep{2014SPIE.9150E..15D} to draw a sequence of r-band observations from the \textit{minion\_1016} simulated survey \footnote{See the list of simulated surveys at \url{https://www.lsst.org/scientists/simulations/opsim/opsim-survey-data}.}. The observations cover a broad range of positions and observing conditions.

In the second stage, we use the LSST Catalog Simulator \citep{2014SPIE.9150E..14C} to produce a full density and LSST-depth (dense) catalog for each observation. We clip the catalog in a .24 square degree box that roughly corresponds to a single LSST CCD. Each catalog contains around 50,000 sources. 

The third stage of simulation is the most important. We use the LSST Photon Simulator (\textit{PhoSim}, \citealt{2014JInst...9C4010P, 2015ApJS..218...14P}) to map the simulated catalogs into high fidelity simulated LSST images. PhoSim uses monte carlo sampling to draw photons from astronomical sources and employs a variety of physics codes to simulate the propagation of photons and electrons through the atmosphere, telescope, and camera. PhoSim has been optimized to make these rich, and expensive simulations tractable. While PhoSim simulations are more computationally expensive than other alternatives, their accuracy is superior and gives more credibility to our results.

The treatment of the atmosphere in PhoSim is particularly important for this high time resolution application. PhoSim employs a raytrace approximation that separates the refractive and diffractive contributions from atmospheric turbulence \citep{2015ApJS..218...14P}. While this approximation does not properly reproduce full speckle patterns, it does accurately reproduce the image jitter in the direction perpendicular to the trail and the intensity modulation along the trail.

The length of star trails can be computed with a simple formula. The LSST pixel pitch and exposure time are 0.21 arcseconds and 15 seconds. Combining this with the roughly 15 arcseconds/second rotation of the Earth we have that, for a given declination $\delta$, the star trail length is $\sim 3.75\cdot\cos(\delta)$ arcminutes, or $ \sim 1071\cdot\cos(\delta)$ pixels. As the declination transitions from the equator to either of the poles, the curvature in the trails increases and their length decreases. If the LSST rotates with the Earth it would take $\sim 47 / \cos(\delta)$ milliseconds for a stellar point source to cross the 0.7 arcsecond LSST PSF. While we cannot resolve separate events below this temporal limit, it is possible to \textit{detect} single events on even shorter time scales.

We use bursts, a brief period when a star's intrinsic flux is increased, as a proxy for stellar variability. They correspond to a tophat in the light curve and are parameterized by their start time, duration, and the source magnitude change. We augment the PhoSim input catalog scheme with these variable parameters and create new simulation modes to produce our data\footnote{Our modifications to PhoSim can be found on Bitbucket: \url{https://bitbucket.org/davidthomas5412/phosim_release}.}. The modes are described below and displayed in Figure \ref{fig:1}.
\begin{itemize}
\item Static: There are no changes to PhoSim in this mode. It produces standard static images.
\item Non-variable: This mode produces trail images but does not simulate the bursts.
\item Variable: This mode produces trail images and simulates the bursts.
\item Burst-only: This mode produces images that only contain the photons from the burst that are in excess of the star's steady flux. 
\end{itemize}
In theory, the burst-only images are equivalent to subtracting the photon counts in the non-variable image from the variable image. In practice, the non-variable and variable images have differences stemming from different random generator access patterns due to the burst photons. We create the burst-only mode to fix this discrepancy and match exactly, pixel by pixel, photon by photon, with the bursts in the variable image. Because these images serve as the labels for training our deep neural network, this pixel level consistency is crucial.

\begin{figure*}[htb]
\center
\includegraphics{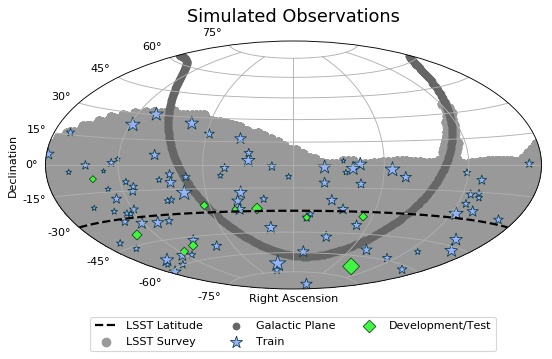}
\caption{The 100 LSST r-band observations used in our experiment. The blue stars are observations in the training set; the green diamonds are observations in the development and test sets. The size of the marker is proportional to the seeing on the evening of the observation. The dotted black line is the latitude of the LSST, the light gray area is the \textit{minion\_1016} LSST survey coverage, and the dark grey line is the galactic plane.}
\label{fig:2}
\end{figure*}

Training, validating, and testing our deep neural network requires many simulated images. We start with 100 r-band observations (shown in Figure \ref{fig:2}). For each observation we produce one dense \textit{background} catalog and ten sparse \textit{foreground} catalogs with 100 variable sources each. Strategically combining each computationally expensive background image with its corresponding 10 foreground images allows us to generate ten times more images with a comparable amount of computation. 

The variable foreground sources have AB magnitudes, burst magnitude changes, and burst durations that are drawn from uniform distributions ranges that are motivated as follows. The source AB magnitude range is 14 to 20. Brighter sources begin taking over a day to simulate in a single thread while fainter star trails are difficult to discern from typical sky backgrounds. The burst magnitude change range is -2.50 (10x flux) to -0.75 (2x flux). This high range provides a clean separation between intrinsic flux changes and the Poisson noise in the star trail. The burst duration range is 10 milliseconds to 1 second. It is difficult to resolve below 10 milliseconds considering the aforementioned $\sim 47 / \cos(\delta)$ millisecond time to trail through a typical LSST PSF. On the other hand, bursts lasting longer than 1 second can potentially be detected with conventional CCDs.

The backgrounds contain no variable sources and we only simulate them in non-variable mode. We simulate the foregrounds in non-variable, variable, and burst-only modes but with the background noise turned off. We add the non-variable and variable foregrounds to their corresponding backgrounds. We are left with 1,000 full density non-variable, variable, and burst-only images.

Each training sample consists of an input and a label image. We use a sequence of transformations to generate many samples from our initial non-variable, variable, and burst-only images. We take the log of the pixel values, random $512 \times 512$ pixel crops, random rotations, and down-sample the image by a factor of two by averaging along both axes. This produces a $256 \times 256$ pixel input image. We apply this procedure to either a pair of corresponding variable and burst-only images or a pair of non-variable and empty images. Half of the training samples are non-variable; the other half are variable.

We split our data into training, development, and test sets. The training set is used to train the network. The development set is used to monitor the network during training and measure its ability to generalize to samples it was not trained on. We also use the development set to make architecture and training process optimizations. The test set is used for a final blind evaluation and to confirm we have not overfit the development set through our research process. In order to ensure that the images in the development and test sets are completely separate from the data the network is trained on, we split the 100 original backgrounds into 90 that are used generate the training set and 10 that are used to generate the development and test sets. This partition is shown in Figure \ref{fig:2}. Then we generate 180,000 training samples for the training set, and 100 each for the development and test sets.

\section{Network and Training}
\label{sec:network}

\begin{figure*}[htb]
\center
\includegraphics[width=\textwidth]{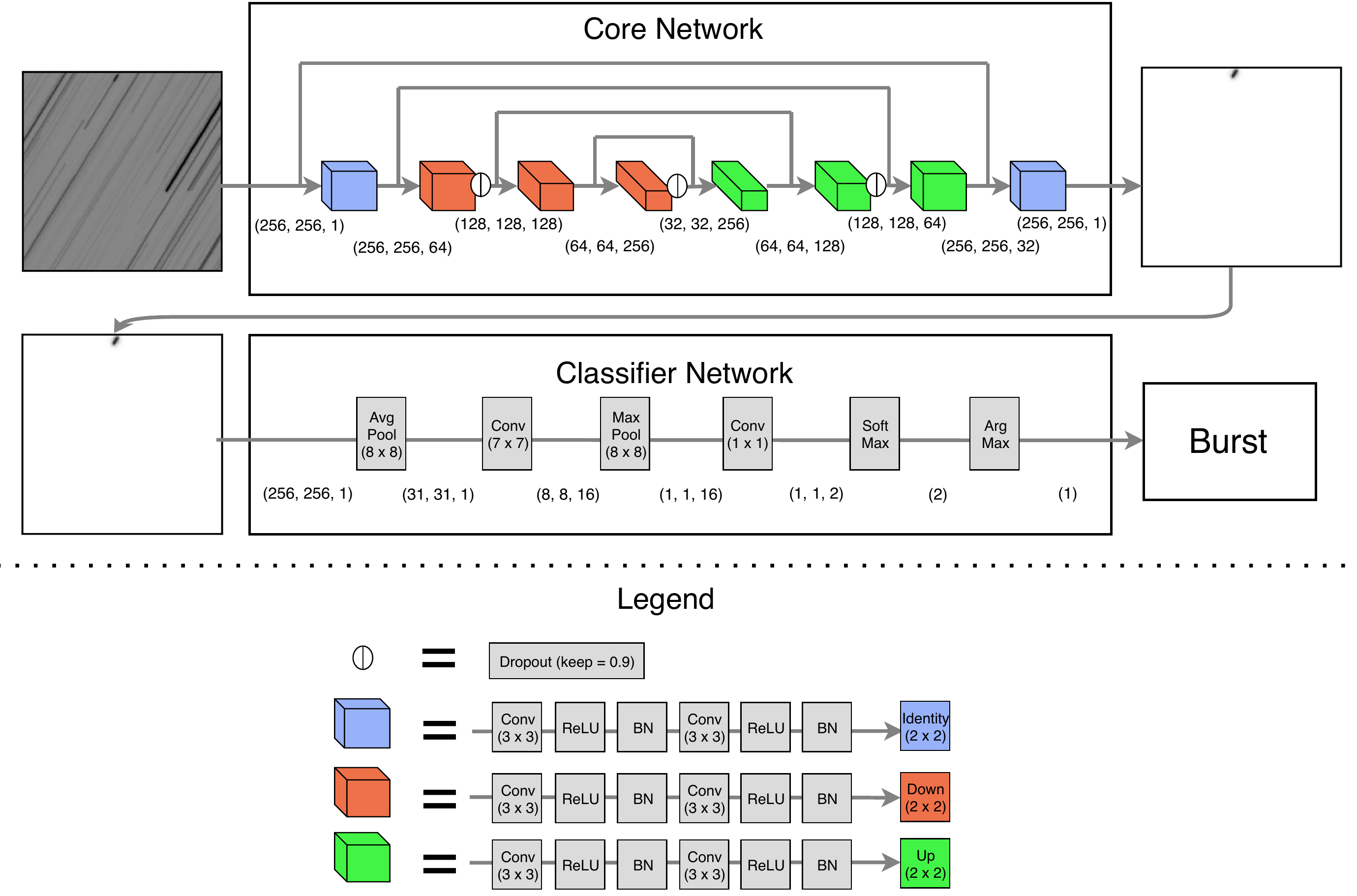}
\caption{The core and classifier network architectures. The gray lines and arrows show the path of tensors in the forward pass of the networks from the input images to the output. The adjacent tuples show the dimensions of the tensors. The blue, red, and green boxes in the core network represent different blocks of layers. The legend at the bottom shows the symbol for dropout and the specific layers in each block.}
\label{fig:3}
\end{figure*}

The goal of our network is to find bursts in star trail images. We solve this task in two stages and train a separate neural network for each. First, we train a \textit{core} network to find the burst flux. Second, we train a \textit{classifier} network to determine whether the burst flux represents a true detection. Figure \ref{fig:3} summarizes the process and our architecture.

The first network takes a simulated LSST image as input, and tries to produce the burst flux label image as output. This problem is similar to the problem of image segmentation. In both tasks the network is trained to find regions of interest in the input and propagate most of their structure to the output. Common features of these convolutional neural networks include a down-sampling phase followed by an up-sampling phase, no fully connected layers, narrow receptive fields, many skip connections, and general simplicity.

Our architecture is inspired by the U-Net and SegNet architectures \citep{2015arXiv150504597R,badrinarayanan2015segnet2}, but has some important differences. Instead of using a Softmax layer to generate different segmentation classes, we predict the raw photon counts from excess burst flux. We use the pixel-wise L2 norm between the network output and burst image label for the cost function. This cost function encourages the network to learn to predict the flux in each pixel during the training process. The steeper L2 norm leads to better performance in this task than the L1 norm or weighted L1 norm.

Each convolution layer is followed by a rectified linear unit (ReLU) activation and batch normalization (BN). Networks with many layers are susceptible to covariate shift, which is when updates to one layer of the network change the distribution of inputs to another layer. The BN layers mitigate covariate shift and allow the network to learn faster \citep{Ioffe:2015:BNA:3045118.3045167}. We employ three dropout layers, which randomly remove nodes throughout the training process, to prevent the network from relying too heavily on specific nodes and overfitting \citep{JMLR:v15:srivastava14a}. Our chosen network dimensions balance performance and complexity. For example, having fewer layers degrades performance and produces lower scores in training while having more layers does not noticeably improve the scores. 

We train the core network for 5 epochs over the 180,000 training set pairs and update the network weights with the ADAM optimizer \citep{journals/corr/KingmaB14}. We use a batch size of 16 to smoothen the gradient descent and to eliminate covariate shift via the batch normalization layers. Throughout training we regularly evaluate the network on the development set to track its performance. After training is complete, we confirm that the network's score on the development set is close to the network's score on the test set.

The classification network takes the burst flux images output by the core network and produces the probability that there was a burst in the original image. We find that a simple neural network achieves superior performance to fixed photon count cutoffs. The average pool and max pool layers encourage the network to focus on the region with the highest flux concentration.

We train the classification network for 5 epochs over 10,000 training samples, in batches of 16. The training samples are the output from the core network on a subset of its training set. Again, once training is complete, we confirm that the network's score on the development set is similar to the network's score on the test set. 

We use the Pytorch Python package to implement and train both networks \citep{paszke2017automatic}. The parameters of the core and classifier networks have 13 MB and 7.2 KB footprints respectively.

\section{Results}
\label{sec:results}

\begin{figure*}[htb]
\center
\includegraphics[width=\textwidth]{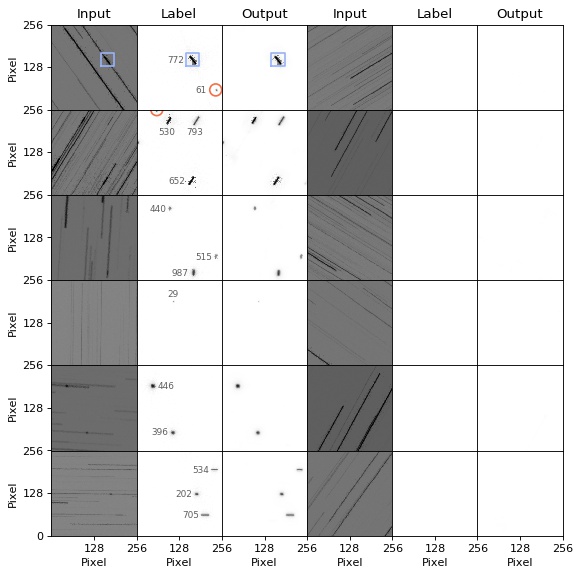}
\caption{The inputs, labels, and outputs for twelve representative test samples. The six samples on the left contain bursts, which are annotated with their duration in milliseconds. The six samples on the right do not contain bursts. The blue boxes in the upper left images are examined in more detail in Figure \ref{fig:5}. The red circles highlight bursts that the network fails to detect.}
\label{fig:4}
\end{figure*}

We characterize the performance of our network with qualitative and quantitative results. Figure \ref{fig:4} shows the output of our core network on twelve test samples. The input images have different background levels, point spread functions (PSFs), flux order of magnitudes, star trail orientations, and star trail lengths and curvature. We see that the network is capable of extracting bursts in a wide variety of situations.

Deep neural networks are parameterized by many weights (3,296,289 for our core network) which makes them challenging to interpret. One way to get insight into the network is to study the cases it fails on. The first two samples in the upper right of Figure \ref{fig:4} show representative failure cases. In both examples, the core network fails to pick out a faint burst amid multiple bursts. The core network fails on examples that are brief, faint, or ambiguous. These examples are intrinsically more difficult because they have lower signal to noise. Similarly, the field of view might only contain a small segment of a trail, which can make burst extraction fundamentally ambiguous in the sense that either prediction - burst flux or no burst flux - is physically plausible.

\begin{figure*}[htb]
\center
\includegraphics{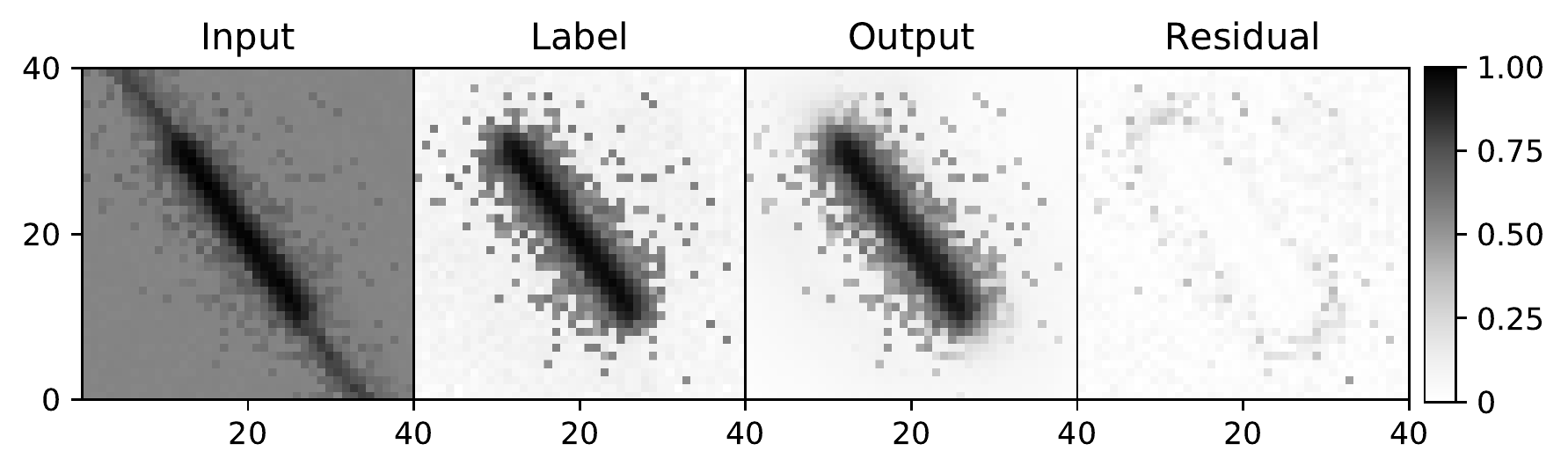}
\caption{Zoom-in images of the burst highlighted in Figure \ref{fig:4}. The residual image is the absolute difference between the label and output images.}
\label{fig:5}
\end{figure*}

Figure \ref{fig:5} shows zoomed-in images of a detected burst. The network successfully extracts individual pixels with excess flux. It even recognizes photons that have scattered away from the star trail. The residual pixel values are a few orders of magnitude less than the photon counts in the burst. The large signal to noise ratio suggests that further scientific analysis can be done directly on the network output.

\begin{figure}[htb]
\center
\includegraphics[width=\columnwidth]{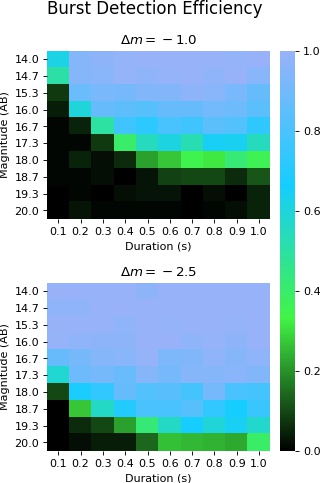}
\caption{The burst detection efficiency heat maps of the combined network on never before seen variable images. Each test sample contains one burst with a specified source magnitude and burst duration drawn from the $10 \times 10$ grid. The plots on the top and bottom correspond to sources with burst magnitude changes of -1.0 and -2.5 respectively.}
\label{fig:6}
\end{figure}

We assess the performance of the combined core and classifier network on a variety of inputs. If the probability of a burst output by the network is greater than its complement, we consider it a burst classification. The false-positive rate on new non-variable samples is 0.86\%, effectively negligible. Thus we focus on the burst detection rate on variable samples, which we deem the efficiency. We make a grid of 10 source AB magnitudes by 10 burst durations. For each grid point we generate 100 new foreground test images that each contain a single burst with the given parameters. We add these foregrounds to the corresponding backgrounds. Then we evaluate the combined network on the samples from each parameter grid point and compute the efficiency. We do this for bursts with magnitude changes of -1.0 and -2.5. The resulting heat maps are shown in Figure \ref{fig:6}.

The efficiency is close to 100\% for a large portion of the parameter space. As expected, it decreases on higher source magnitudes and shorter burst durations. A larger burst magnitude change expands the efficiency out to higher source magnitudes. The network can detect 1 second bursts out to 20th magnitude when the burst magnitude change is -2.5.

\begin{figure*}[htb]
\center
\includegraphics{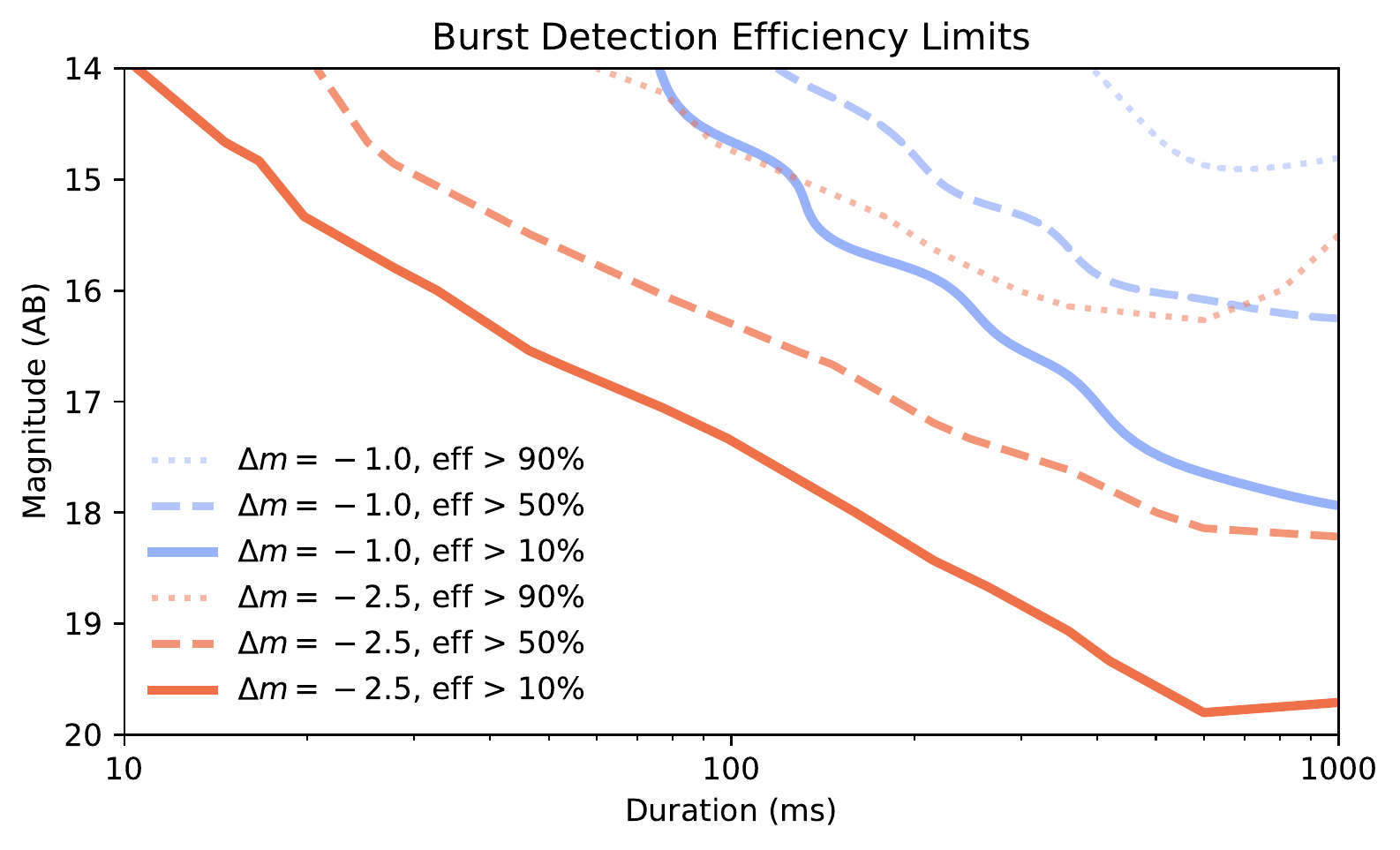}
\caption{The contours show the probability the combined network will detect a burst with the given source magnitude and burst duration. The blue and red contours correspond to burst magnitude changes of -1.0 and -2.5 respectively. The dotted, dashed, and solid lines correspond to 90\%, 50\% and 10\% detection efficiencies respectively.}
\label{fig:7}
\end{figure*}

We also test the time resolution of our method by repeating the above experiment with a new grid of 10 logarithmically spaced burst durations down to 10 milliseconds. The contour plots in Figure \ref{fig:7} show how the combined network performs on short time scales. The network has 69\% efficiency on 14th magnitude bursts lasting 25 milliseconds with a -2.5 magnitude change. This time resolution is competitive with existing high time resolution instruments with wide-field imaging \citep{2016MNRAS.457.3036H}. 

The sources in the training set have burst durations that are sampled uniformly from 10 milliseconds to 1 second. This means that very few training samples are in the tens of milliseconds. In expectation, approximately 1 in 100 bursts in the training images have duration less than 20 milliseconds. Increasing the proportion of very short duration bursts in our training set may improve the time resolution of the network even further.

\begin{figure*}[htb]
\center
\includegraphics[width=0.95\textwidth]{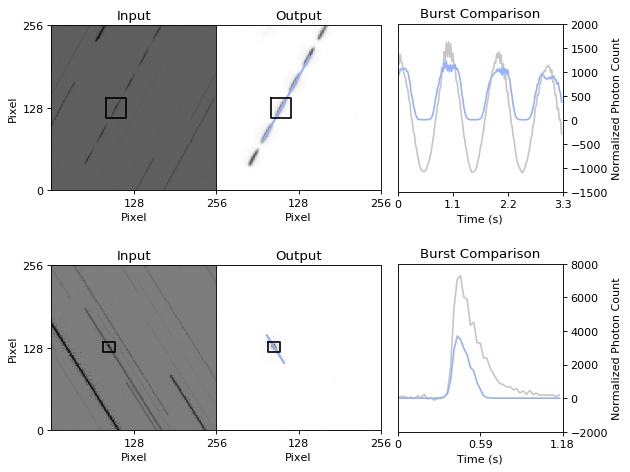}
\caption{The inputs, outputs, and burst comparisons for the sinusoidal (top) and exponentially decaying (bottom) samples. The photon counts in the comparisons come from fitting a line to the burst, shown in blue, and linearly interpolating the photon counts along it in both the input and output images. We normalize the input photon counts by subtracting the median photon count along the trail. 
}
\label{fig:8}
\end{figure*}

The far-reaching success of deep learning suggests that these networks are not memorizing examples but fundamentally learning how to solve tasks and generalize to new samples. Leading AI researchers have proposed explanations \citep{2016arXiv161103530Z,2017arXiv170605394A}. We perform an experiment to examine the extent to which our network can generalize to new yet related problems. We simulate two stars with new variability curves. The first oscillates sinusoidally. The second star's flux increases as a step function and decays exponentially. These are both different than the bursts our network was trained on, but similar in that they both involve extreme flux variations. After evaluating our core network on these new input images, we compare the interpolated photon counts along lines through the relevant trail segments. Figure \ref{fig:8} shows our findings. In both cases, our network picks up the new variations. Not only does this confirm that our network is learning a general task, it also hints that we might be able to train a network to recognize arbitrary variability with a simple basis, such as tophats, or bursts.

\section{Discussion}
\label{sec:discussion}
The next decade of astronomy will be paced by large projects and remarkable new data volumes. The LSST will capture over 3 billion new pixels of information every 17 seconds of operation. For algorithms to be applied in this context, not only must they be correct, they also must be scalable. In deep learning, the training is computationally expensive. It takes close to 6 hours to train our network with an Intel Xeon E5-2640 v4 CPU and NVIDIA Tesla P100 GPU connected via PCIe. However once this one-time cost to train the network is paid, the network can evaluate new samples rapidly. It takes 2.8 and 0.3 milliseconds to evaluate a sample with the core and classifier networks respectively. We measured this by evaluating 1,000 different samples and taking the difference from the time the data is in CPU memory until the output is in CPU memory. In order to scale our method to an entire LSST image, we would divide each of the 189 CCD images into 64 $512 \times 512$ patches (12,096 patches in total), which would take 37 seconds to evaluate sequentially. Given that each patch can be evaluated independently, we can make further performance gains with parallelization. The speed and scalability are notable advantages of our approach.

Another strength of our method is its simplicity. Deep learning allows us to formulate the burst detection problem in a straightforward manner. The only input is the star trail image. We do not include information about the specific sources, PSF, star trail curvature, telescope pointing, or observing conditions. The network effectively designs its own features in the training process. It then employs these optimized features to extract bursts in images and classify them correctly. This process is simple and requires few assumptions. Moreover, the resulting network is robust and easy to deploy.

Deep learning typically requires large training sets. In this work, we employ simulations to produce the obligatory data volume. This allows us to control many aspects of the problem. To change how the network performs on different classes of images we simply manipulate the composition of the training set. We can generate new simulated training data on the fly. We can tailor the data to teach the network specific functionality. We can measure the accuracy of our network to arbitrary empirical precision. Deep learning combined with simulation is an incredibly advantageous workflow.

We present two other potential approaches to detecting photometric outliers in wide field star trail images. One alternative approach would be to extended conventional image subtraction, such as the popular ZOGY algorithm \citep{2016ApJ...830...27Z}, to subtract star trail images from static reference images. There are a few challenges with this approach. The flux from a source is spread out over many pixels in a star trail image, which must be taken into account when handling saturation and bleeding in the subtraction. There are also two sources of contamination. There is contamination from sources which are initially outside the field of view, but trail in during the exposure. There is also, depending on the convolution method, contamination from sources that are initially inside the field of view, but trail out during the exposure. When the static image sources are convolved with a trail kernel via a fourier transform, the periodic assumptions underlying the fourier transform lead to contamination on the other side of the image. The quadratic runtime of employing an explicit, non-fourier-transform-based convolution is prohibitive.

These boundary contaminations encourage one to use image sizes that are as large as possible, at least larger than the equatorial plane star trail length of 3.75 arcminutes, or 1071 pixels. However, slight PSF variations across the star trails cause wiggles in the trails that while correlated on small angular scales, below approximately 1 arcminute, are increasingly uncorrelated on the scales needed to avoid significant image contamination. This transverse wiggling produces artifacts that are challenging to collectively eliminate. Furthermore, finding bursts through image subtraction typically requires more dependencies such as reference images, fitted PSFs, and various uncertainties. Thus even if an algorithm could address the aforementioned challenges, it would likely require more dependencies than our solution. 

Another approach is inspired by slitless spectroscopy. The idea would be to perform spectrograph extraction as in \citealt{2009PASP..121...59K} or \citealt{2016ApJS..225...27M} but replace spectrums with star trails. These methods typically rely on a static image to find sources and map them to spectra in the spectral image. The spectra, or star trails in our case, are then analyzed individually. There may be use cases where this granular analysis is preferred to our method of employing a neural network to dramatically filter the number of sources of interest before applying more thorough analysis.

There are many avenues for future work. A major assumption in our work is that a network trained on simulated images will generalize to real images. It is important to confirm this assumption with real data and find ways to potentially incorporate additional effects into the simulations. We are planning to take star trail images with a wide-field telescope and further validate our method.

There is a wide range of science that can benefit from an instrument that combines a wide field of view with the capability of scanning for short duration stellar variability. While a network trained to detect general variability may serve many use cases, it is conceivable that having a suite of dedicated networks, optimized for their specific use cases, could enhance performance further.

There are also degrees of freedom inherent to the problem that may be fruitful to optimize. The telescope motion and exposure time are two such interesting variables. Through ``anti-tracking'' for example, we can produce longer star trails, which would enable higher time resolution. However, the flux from a given source will be increasingly spread out over the detector, degrading our ability to detect faint sources. A similar tradeoff exists between exposure time and the number of star trail intersections. 

The false-positive rate for our network is less than 1\%, a minute fraction. But the number of false-positive detections grows linearly with the imaging workflow. For extremely large workflows, with few true events, the false detections can inundate follow-up efforts and stagnate the discovery process. One way to decrease the rate of false-positive follow-ups is to make multiple evaluations on slightly different images that all contain the region of interest. We can arbitrarily rotate the image or dither the $512 \times 512$ pixel crop from the larger image to get multiple evaluations. The consistency and consensus in the suite of predictions can be used to filter the detections chosen for follow-up.

There is ample literature demonstrating the susceptibility of deep neural networks to adversarial examples and perturbations that can be imperceptible to humans \citep{2014arXiv1412.6572G}. While this is a concern in many production environments, it can also be used to better understand how a neural network fails. Taking the gradient of the cost function, with respect to the input pixels, while keeping the weights fixed, gives the  perturbation to a particular input that the network is most sensitive to. Any patterns across different input samples can be used to reinforce the training process and further enhance the robustness of the network.

Detecting variability is the first step in a larger data reduction. After applying our method, scientists can follow up with a more sophisticated and computationally expensive photometric extraction and parameter inference on the sources of interest. This will likely involve computing the Bayesian evidences of competing hypotheses, such as burst or no burst, and doing Bayesian inference with Markov Chain Monte Carlo methods to produce the posterior distributions for the event parameters. We look forward to stimulating these more conventional analyses with the events detected by our network.

\section{Conclusions}
\label{sec:conclusion}

To conclude, we summarize our main results as follows.
\begin{enumerate}
\item We train a deep neural network to detect bursts in simulated LSST star trail images. 
\item We demonstrate that our network is robust. The bursts the network fails to detect are intrinsically difficult cases. The network performs well on the test set and it generalizes to new types of variability.
\item We empirically confirm that star trails enable sub-exposure time resolution. The combined network has less than 1\% false-positives, detects variability in sources out to 20th magnitude, and detects bursts on time scales down to 10 milliseconds. 
\end{enumerate}

The primary implication of this work is that by taking star tail images and processing them with deep learning we can extend the scientific payload of large survey telescopes to the vast field of high time resolution astrophysics. A number of astrophysics communities can benefit from such observations. 

\acknowledgments

This material is based upon work supported in part by the National Science Foundation through
Cooperative Agreement 1258333 managed by the Association of Universities for Research in Astronomy
(AURA), and the Department of Energy under Contract No. DE-AC02-76SF00515 with the SLAC National
Accelerator Laboratory. Additional LSST funding comes from private donations, grants to universities, and in-kind support from LSSTC Institutional Members.

Some of the computing for this project was performed on the Sherlock cluster. We thank Stanford University and the Stanford Research Computing Center for providing computational resources and support that contributed to these research results. This research made use of the Python Programming Language, along with many community-developed or maintained software packages, including
Astropy \citep{astropy},
IPython \citep{ipython},
Jupyter (\http{jupyter.org}),
Matplotlib \citep{matplotlib},
NumPy \citep{numpy},
Pandas \citep{pandas},
PyTorch \citep{paszke2017automatic},
and SciPy \citep{scipy}.
This research made extensive use of the {\tt arXiv} and NASA's Astrophysics Data System for bibliographic information. David Thomas thanks the LSSTC Data Science Fellowship Program, his time as a Fellow has benefited this work.

\bibliographystyle{aasjournal}
\bibliography{../references}
\end{document}